\documentclass[10pt]{IEEEtran} 

\usepackage{graphicx}         
\usepackage[latin1]{inputenc} 

\usepackage{amssymb}          
\usepackage{amsmath}          
\usepackage{amsfonts}

\newtheorem{pro}{Proposition}
\newtheorem{thm}{Theorem}
\newtheorem{cor}{Corollary}
\newtheorem{lem}{Lemma}

\newtheorem{dfn}{Definition}

\newcommand{\ket}[1]{|#1\rangle}
\newcommand{\bra}[1]{\langle #1|}

\newcommand{\Hi}{\mathcal{H}}

\newcommand{\Tr}{\mathrm{{Tr}}}
\newcommand{\supp}{\textrm{supp}}

\newcommand{\C}{\mathbb{C}}

\newcommand{\trace}{\textrm{trace}}

\def\smallfrac#1#2{{\textstyle\frac{#1}{#2}}}

\newcommand{\beq}{\begin{equation}}
\newcommand{\eeq}{\end{equation}}
\newcommand{\beqa}{\begin{eqnarray}}
\newcommand{\eeqa}{\end{eqnarray}}
\newcommand{\beqan}{\begin{eqnarray*}}
\newcommand{\eeqan}{\end{eqnarray*}}
\newcommand{\bea}{\begin{eqnarray}}
\newcommand{\eea}{\end{eqnarray}}

\newcommand\bm[4]{
	\begin{bmatrix}
		#1 & #2\\
	    #3 & #4
	\end{bmatrix}
}

\newcommand\bml[4]{
	\left[\begin{array}{c|c}
		#1 & #2\\
		\hline 
	    #3 & #4
	\end{array}\right]
}

\newcommand\M{ M_{k}^{\phantom{\dagger}}}				
\newcommand\Ma{	M_{k}^{\dagger}}						
\newcommand\Mb[1]{ M_{k,#1}^{\phantom{\dagger}}}		
\newcommand\Mab[1]{	M_{k,#1}^{\dagger}}					

\newcommand\mat[1]{ #1_{k}^{\phantom{\dagger}}}			
\newcommand\mata[1]{	#1_{k}^{\dagger}}					

\title{Engineering Stable Discrete-Time Quantum Dynamics via a Canonical QR Decomposition}

\author{Saverio~Bolognani and~Francesco~Ticozzi%
\thanks{S. Bolognani is with the Dipartimento di Ingegneria dell'Informazione, Universit\`a di Padova, via Gradenigo 6/B, 35131 Padova, Italy ({\tt saverio.bolognani@dei.unipd.it}).}%
\thanks{F. Ticozzi is with the Dipartimento di Ingegneria dell'Informazione, Universit\`a di Padova, via Gradenigo 6/B, 35131 Padova, Italy ({\tt ticozzi@dei.unipd.it}).}%
\thanks{Work partially supported by the CPDA080209/08 research grant of the University of Padova, and by the Department of Information Engineering research project ``QUINTET''.}%
}

\date{\today}

\begin{document}
\maketitle

\begin{abstract} 
We analyze the asymptotic behavior of discrete-time, Markovian quantum systems with respect to a subspace of interest.
Global asymptotic stability of subspaces is relevant to quantum information processing,
in particular for initializing the system in pure states or subspace codes.
We provide a linear-algebraic characterization of the dynamical properties leading to invariance and attractivity of a given quantum subspace.
We then construct a design algorithm for discrete-time feedback control that allows to stabilize a target subspace, proving that if the control problem is feasible, then the algorithm returns an effective control choice. 
In order to prove this result, a canonical QR matrix decomposition is derived,
and also used to establish the control scheme potential for the simulation of open-system dynamics.
\end{abstract}

\begin{IEEEkeywords} Quantum control, QR decomposition, invariance principle, quantum information.
\end{IEEEkeywords}

\section{Introduction}

\IEEEPARstart{S}{ince} the pioneering intuitions of R. P. Feynmann \cite{feynman-QC}, Quantum Information (QI) has been the focus of an impressive research effort. Its potential has been clearly demonstrated, not only as a new paradigm for fundamental physics, but also as the key ingredient for a new generation of information technologies. Today the goal is to design and produce quantum chips, quantum memories, and quantum secure communication protocols \cite{expreview,nielsen-chuang,zeilinger}. 
The main difficulties in building effective QI processing devices are mainly related to scalability issues and to the disruptive action of the environment on the quantum correlations that embody the key advantage of QI. Many of these issues do not appear to be fundamental, and their solution is becoming mainly an engineering problem.

Most of the proposed approaches to realize quantum information technology require the ability to perform sequences of a limited number of fundamental operations. Two typical key tasks are concerned with the preparation of states of maximal information \cite{divincenzo,nielsen-chuang,viola-qubit} and engineering of protected realization of quantum information \cite{viola-dd,lidar-dfs,viola-generalnoise,knill-protected}, i.e. the realization of information encodings that preserve the fragile quantum states from the action of noise. This paper will focus on these issues, providing a design strategy for engineering stable quantum subspaces.


In \cite{shabani-lidar} and \cite{ticozzi-QDS} the seminal linear-algebraic approach of \cite{lidar-dfs} to study noise-free subspaces has been extended to the general setting of noiseless subsystems \cite{viola-generalnoise} (which usually entails an operator-algebraic approach) and developed in two different directions, both concerned with the {\em robustness} of the encoded quantum information. The first (\cite{shabani-lidar}) studies the cases in which the encoded information does not degenerate in the presence of initialization errors, the other (\cite{ticozzi-QDS}) aims to ensure that the chosen encoding is an invariant, asymptotically stable set for the dynamics in presence of the noise. The latter tightly connects the encoding task to a set of familiar {\em stabilization} control problems. 

Feedback state stabilization, and in particular pure-state stabilization problems, have been tackled in the quantum domain under a variety of modeling and control assumptions, with a rapidly growing body of work dealing with the Lyapunov approach, see e.g. \cite{grivopoulos, altafini-feedback, wiseman-feedback, vanhandel-feedback, raccanelli-cdc, mirrahimi-lyapunov, ticozzi-QDS,ticozzi-markovian,schirmer-1,schirmer-2,dalessandro-book} and references therein.
Here we embrace the approach of \cite{ticozzi-QDS, ticozzi-markovian}, extending these results to Markovian discrete-time evolutions. 

A good review of the role of discrete-time models for quantum dynamics and control problems can be found in \cite{vanhandel-invitation}, to which we refer for a discussion of the relevant literature which is beyond  the scope of this paper.
In fact, we will assume from the very beginning discrete-time quantum dynamics described by sequences of trace-preserving quantum operations in Kraus representation \cite{kraus,nielsen-chuang}. This assumption implies the Markovian character of the evolution \cite{kummerer-markov}, which, along with a forward composition law, ensures a semigroup structure. We introduce the class of dynamics of interest and the relevant notation in Section \ref{section-qds}. A basic analysis of kinematic controllability for Kraus map has been provided in \cite{kraus-controllability}.

After introducing the key concepts relative to {\em quantum subspaces} and dynamical stability in Section \ref{qsubspaces}, Section \ref{gas} is devoted to the analysis of the dynamics. The results provide us with necessary and sufficient conditions on the dynamical model that ensure global stability of a certain quantum subspace. We employ a Lyapunov approach, exploiting the linearity of the dynamics, as well as the convex character of the state manifold. Lyapunov analysis of quantum discrete-time semigroups has been also considered, with emphasis on ergodicity properties, in \cite{giovannetti-lyapunov}. 

The control scheme we next consider modifies the underlying dynamics of the system by indirectly measuring it, and applying unitary control actions, conditioned on the outcome of the measurement. If {\em we average over the possible outcomes}, we obtain a new semigroup evolution where the choice of the control can be used to achieve the desired stabilization. 
We make use of the generalized measurement formalism, which is briefly recalled in Appendix \ref{measurements}. This control scheme can be seen as an instance of discrete-time {\em  Markovian
reservoir engineering}: the use of ``noisy'' dynamics to obtain a desired dynamical behavior has long been investigated in a variety of contexts, see {\em
e.g.}  \cite{poyatos,davidovich,dematos,cirac-engineering}.

The synthesis results of Section \ref{engineering} include a simple characterization of the controlled dynamics that can be enacted, and an algorithm that builds unitary control actions stabilizing the desired subspace. If such controls cannot be found, it is proven that no choice of controls can achieve the control task for the same measurement.  The main tools we employ come from the stability theory of dynamical systems, namely LaSalle's Invariance principle \cite{lasalle-discrete}, and linear algebra, namely the QR matrix decomposition \cite{horn-johnson}. We shall construct a ``special form'' of the QR decomposition: In particular, we prove that the upper triangular factor $R$ can be rendered a {\em canonical form} with respect to the left action of the unitary matrix group. This result and the related discussion is presented in Section \ref{sec:canonicalForm}.




%


\section{Discrete--time quantum dynamical semigroups}\label{section-qds}

Let $\mathcal I$ denote the physical quantum system of interest. Consider the associated separable Hilbert space $\mathcal H_I$ over the complex field $\mathbb C$.  In what follows, we consider finite-dimensional quantum systems, i.e. $\dim(\mathcal H_I) < \infty$.
In Dirac's notation, vectors are represented by a
{\em ket} $\ket{\psi}\in\Hi_I,$ and linear functionals by a {\em bra},
$\bra{\psi}\in\Hi_I^\dag$ (the adjoint of $\Hi_I$), respectively. The inner product of
$\ket{\psi},\ket{\varphi}$ is then represented as $\langle \psi |
\varphi \rangle.$

Let $\mathfrak B(\mathcal H_I)$ represent the set of linear bounded operators on $\mathcal H_I$, $\mathfrak H(\mathcal H_I)$ denoting the real subspace of hermitian operators, with $\mathbb I$ and $\mathbb O$ being the identity and the zero operator, respectively. Our (possibily uncertain) knowledge of the state of the quantum system is condensed in a density operator, or {\em state} $\rho$, with $\rho \ge 0$ and $\Tr \rho = 1$. Density operators form a convex set $\mathfrak D(\mathcal H_I) \subset \mathfrak H(\mathcal H_I)$, with one-dimensional projectors corresponding to extreme points (pure states, $\rho_{|\psi\rangle} = |\psi\rangle\langle\psi|$). Given an $X\in\mathfrak H(\mathcal H_I),$ we indicate with $\ker (X)$ its kernel ($0$-eigenspace) and with $\supp(X):=\Hi_I\ominus \ker(X)$  its range, or {\em support}.
If a quantum system ${\cal Q}$ is obtained by composition of two {\em subsystems} ${\cal
Q}_1,\,{\cal Q}_2$, the corresponding mathematical description is
carried out in the tensor product space, $\Hi_{12}=\Hi_1\otimes\Hi_2$
\cite{sakurai}, observables and density operators being associated
with Hermitian and positive-semidefinite, normalized operators on
$\Hi_{12}$, respectively. 


In the presence of coupling between subsystems, quantum measurements (see Appendix \ref{measurements}), or interaction with surrounding environment, the dynamics of a quantum system cannot in general be described by Schr\"odinger's dynamics: The evolution is no longer unitary and reversible, and the formalism of open quantum systems is required \cite{davies,petruccione,alicki-lendi,nielsen-chuang}.
An effective tool to describe these dynamical systems, of fundamental interest for QI, is given by quantum operations \cite{nielsen-chuang,kraus}. The most general, linear and physically admissible evolutions which take into account interacting quantum systems and measurements, are described by Completely Positive (CP) maps, that via the Kraus-Stinespring theorem \cite{kraus} admit a representation of the form 
\begin{equation}
\mathcal{T}[\rho] = \sum_k \M \rho \Ma \label{eq:KrausMap}
\end{equation}
(also known as operator-sum representation of $\mathcal T$), where $\rho$ is a density operator and $\{M_k\}$ a family of operators such that the completeness relation
\begin{equation}
\sum_k \Ma \M = I \label{eq:KrausMapConditions}
\end{equation}
is satisfied. Under this assumption the map is then Trace-Preserving and Completely-Positive (TPCP), and hence maps density operators to density operators. We refer the reader to e.g. \cite{alicki-lendi,nielsen-chuang, petruccione, davies} for a detailed discussions of the properties of quantum operations and the physical meaning of the complete-positivity property. 

One can then consider the discrete-time dynamical semigroup, acting on $\mathfrak{D}(\Hi_I),$ induced by iteration of a given TPCP map. The resulting discrete-time quantum system is described by \begin{equation}
\rho(t+1) = \mathcal{T}[\rho(t)] =  \sum_k \M \rho(t) \Ma. \label{eq:DiscreteTimeSystem}
\end{equation}
Given the initial conditions $\rho(0)$ for the system, we can then write
$$
\rho(t) = \mathcal T^t[\rho(0)]\,\quad t=1,2,\ldots
$$
where $\mathcal T^t[\cdot]$ indicates $t$ applications of the TPCP map $\mathcal T[\cdot]$. Hence, the evolution obeys a forward composition law and, in the spirit of \cite{alicki-lendi}, is called a Discrete-time Quantum Dynamical Semigroup (DQDS). Notice that while the dynamic map is linear, the ``state space'' $\mathfrak{D}(\Hi_I)$ is a convex, compact subset of the cone of the positive elements in $\mathfrak{H}(\Hi_I).$

While a TPCP maps can indeed represent general dynamics, assuming dynamics of the form  \eqref{eq:DiscreteTimeSystem}, with $M_k$'s that do not depend on the past states, is equivalent to assume Markovian dynamics (see \cite{kummerer-markov} for a discussion of Markovian properties for quantum evolutions). From a probabilistic viewpoint, if density operators play the role of probability distributions, TPCP maps are the analogue of transition operators for classical Markov chains.


\section{Quantum subspaces, invariance and attractivity}\label{qsubspaces}

In this section we recall some definitions of quantum subspaces invariance and attractivity. We follow the subsystem approach of \cite{ticozzi-QDS, ticozzi-markovian}, focusing on the case of subspaces. This is motivated by the fact that the general subsystem case is derived in the continuous-time case as a specialization with some additional constraints, and that for many applications of interest for the present work, namely {\em pure-state preparation} and {\em engineering of protected quantum information}, the subspace case is enough, as it is suggested by the results in \cite{ticozzi-QDS}.

\begin{dfn}[Quantum subspace]
	A quantum subspace $\mathcal S$ of a system $\mathcal I$ with associated Hilbert space $\mathcal H_I$ is a quantum system
	whose Hilbert space is a subspace $\mathcal H_{S}$ of $\mathcal H_I$,
	\begin{equation}
	  \mathcal H_I = \mathcal H_{S} \oplus \mathcal H_R,                \label{eq:spaceDecomposition}
	\end{equation}
	for some remainder space $\mathcal H_R$.
	The set of linear operators on $\mathcal S$, $\mathcal B(\mathcal H_S)$, is isomorphic to the algebra on $\mathcal H_I$
	with elements of the form $X_I = X_S \oplus \mathbb O_R$.
\end{dfn}

Let $n = \dim(\mathcal H_I)$, $m = \dim(\mathcal H_S)$, and $r = \dim(\mathcal H_R)$, and let $\{|\phi\rangle_j^S\}_{j=1}^m$, $\{|\phi\rangle_k^R\}_{k=1}^r$ denote orthonormal bases for $\mathcal H_S$ and $\mathcal H_R$, respectively. Decomposition \eqref{eq:spaceDecomposition} is then naturally associated with the following basis for $\mathcal H_I$:
$$
 \{|\varphi_l\rangle\} = \{|\phi\rangle_j^S\}_{j=1}^m  \cup  \{|\phi\rangle_k^R\}_{k=1}^r.
$$
This basis induces a block structure for matrices representing operators acting on $\mathcal H_I$:
$$
 X = \bml{X_S}{X_P}{X_Q}{X_R}.
$$
In the rest of the paper the subscripts $S,P,Q$ and $R$ will follow this convention.
Let moreover $\Pi_S$ and $\Pi_R$ be the projection operators over the subspaces $\mathcal H_S$ and $\mathcal H_R$, respectively. The following definitions are independent of the choices of 
$\{|\phi\rangle_j^S\}_{j=1}^m$, $\{|\phi\rangle_k^R\}_{k=1}^r$.

\begin{dfn}[State initialization]
	The system $\mathcal I$ with state $\rho \in \mathcal D(\mathcal H_I)$ is initialized in $\mathcal S$
	with state $\rho_S \in \mathcal D(\mathcal H_S)$ if $\rho$ is of the form
	\beq\label{initstate}
	  \rho = \bml{\rho_S}{0}{0}{0}.
	\eeq
	We will denote with $\mathcal J_S(\mathcal H_I)$ the set of states of the form \eqref{initstate} for some $\rho_S \in \mathcal D(\mathcal H_S)$.
\end{dfn}

\begin{dfn}[Invariance]\label{def:InvariantSubsystem}
	Let $\mathcal I$ evolve under iterations of a TPCP map.
	The subsystem $\mathcal S$ supported on the subspace $\mathcal H_S$ of $\mathcal H_I$ is invariant if the
	evolution of any initialized $\rho \in \mathcal J_S(\mathcal H_I)$ obeys
	$$
	\rho(t) = \bml{\mathcal T_S^t[\rho_S]}{0}{0}{0} \in \mathcal J_S(\mathcal H_I)
	$$
	$\forall t \ge 0$, and with $\mathcal T_S$ being a TPCP map on $\mathcal H_S$.
\end{dfn}

\begin{dfn}[Attractivity]\label{def:AttractiveSubsystem}
	Let $\mathcal I$ evolve under iterations of a TPCP map $\mathcal T$.
	The subsystem $\mathcal S$ supported on the subspace $\mathcal H_S$ of $\mathcal H_I$ is attractive if
	$\forall \rho \in \mathcal D(\mathcal H_I)$ we have:
	$$
	  \lim_{t\rightarrow\infty}\left\|\mathcal T^t(\rho) - \Pi_S \mathcal T^t[\rho] \Pi_S \right\| = 0.
	$$
\end{dfn}

\begin{dfn}[Global asymptotic stability]\label{def:GlobalAsymptoticStability}
	Let $\mathcal I$ evolve under iterations of a TPCP map $\mathcal T$.
	The subsystem $\mathcal S$ supported on the subspace $\mathcal H_S$ of $\mathcal H_I$ is
	{\em Globally Asymptotically Stable} (GAS) if it is invariant and attractive.
\end{dfn}


\section{Analysis Results}\label{gas}

This section is devoted to the derivation of necessary and sufficient conditions on the form of the TPCP map ${\mathcal T}$ for {\em a given} quantum subspace $\mathcal S$ to be GAS. We start by focusing on the invariance property.

\subsection[Invariance of J\_S(H\_I)]{Invariance of $\mathcal J_S(\mathcal H_I)$}

The following proposition gives a sufficient and necessary condition on $\mathcal T$ such that $\mathcal J_S(\mathcal H_I)$ is invariant.

\begin{pro} \label{pro:Invariance}
	Let the TPCP transformation $\mathcal T$ be described by the Kraus map \eqref{eq:KrausMap}.
	Let the matrices $\M$ be expressed in their block form
	$$
	\M = \bm{M_{k,S}}{M_{k,P}}{M_{k,Q}}{M_{k,R}}
	$$
	according to the state space decomposition \eqref{eq:spaceDecomposition}.
	Then the set $\mathcal J_S(\mathcal H_I)$ is invariant if and only if
	\begin{equation}
	\Mb{Q} = 0 \quad \forall k\,. \label{eq:conditionOnMq}
	\end{equation}
\end{pro}
\begin{IEEEproof}
	Verifying Definition \ref{def:InvariantSubsystem} is equivalent to verifying that there exists a TPCP map 
	$\mathcal T_S$ such that
	\begin{equation}
	\mathcal{T}\left[ \bm{\rho_S}{0}{0}{0} \right] = \bm{\mathcal T_S(\rho_S)}{0}{0}{0} \label{eq:InvarianceOfMs}
	\end{equation}
	for all $\rho_S$ in $\mathcal D(\mathcal H_S)$.
	By exploiting the block form of the $\M$ matrices in \eqref{eq:KrausMap} given by the decomposition
	\eqref{eq:spaceDecomposition}, we have
	\begin{equation}
	\begin{split}
	\mathcal T \left[ \bm{\rho_S}{0}{0}{0} \right] &= \sum_k \M \bm{\rho_S}{0}{0}{0} \Ma \\ & = 
	\sum_k \bm{\Mb{S}}{\Mb{P}}{\Mb{Q}}{\Mb{R}} \bm{\rho_S}{0}{0}{0}
	\bm{\Mab{S}}{\Mab{Q}}{\Mab{P}}{\Mb{R}} \\
	&= \sum_k \bm{\Mb{S}\rho_S\Mab{S}}{\Mb{S}\rho_S\Mab{Q}}{\Mb{Q}\rho_S\Mab{S}}{\Mb{Q}\rho_S\Mab{Q}}
	\end{split}
	\label{eq:evolutionForInvariance}
	\end{equation}
	Sufficiency of \eqref{eq:conditionOnMq} to have invariance of $\mathcal J_S(\mathcal H_I)$ is trivial.
	Necessity is given by the fact that the lower right blocks $\Mb{Q}\rho_S\Mab{Q}$
	are positive semi-definite for all $k$'s, and therefore, for \eqref{eq:InvarianceOfMs} to hold, it has to be $\Mb{Q}\rho_S\Mab{Q} = 0\; \forall k$.
	For $\Mb{Q}\rho_S\Mab{Q}$ to be zero for any state $\rho_S \in \mathcal D(\mathcal H_S)$,
	it has then to be $\Mb{Q}=0$. Equation \eqref{eq:InvarianceOfMs} then implies that the completely-positive transformation
$$
\mathcal T_S[\rho_S] = \sum_k \Mb{S} \rho_S \Mab{S}
$$
is also trace preserving.
\end{IEEEproof}


\subsection{Global asymptotic stability of $\mathcal J_S(\mathcal H_I)$}

The main tool we are going to use in deriving a characterization of TPCP maps that render a certain $\Hi_S$ GAS, is LaSalle's invariance principle, which we recall here in its discrete time form \cite{lasalle-discrete}.

\begin{thm}[La Salle's theorem for discrete-time systems] \label{th:LaSalle}
	Consider a discrete-time system
	$$
	x(t+1) = \mathcal{T}[x(t)]
	$$
	Suppose $V$ is a $\mathcal{C}^1$ function of $x\in \mathbb R^n$, bounded below and satisfying
	\begin{equation}
	\Delta V(x) = V(\mathcal{T}[x]) - V(x) \le 0, \quad \forall x \label{eq:LaSalleHypotesis}
	\end{equation}
	i.e. $V(x)$ is non-increasing along forward trajectories of the plant dynamics.
	Then any bounded trajectory converges to the largest invariant subset $W$ contained
	in the locus $E=\{x | \Delta V(x) = 0\}$.
\end{thm}

Being any TPCP map a map from the compact set of density operators to itself, any trajectory is bounded. Let's then consider the function
\begin{equation}
V(\rho)  = \Tr(\Pi_R \rho) \ge 0. \label{eq:LyapFunction}
\end{equation}
The function $V(\rho)$ is $\mathcal C^1$ and bounded from below, and it is a natural candidate for a Lyapunov function for the system. In fact, it represents the probability of the event $\Pi_R$ (see Appendix \ref{measurements}), that is, the probability that the system is found in ${\cal H}_R$ after the measurement. 

\begin{lem} 
Let ${\mathcal T}$ be the generator of a DQDS, and assume a given quantum subsystems $\mathcal S$ to be invariant. Then $V(\rho)  = \Tr(\Pi_R \rho)$ ($\mathcal H_R$ being the remainder space) satisfies the hypothesis \eqref{eq:LaSalleHypotesis} of Theorem \ref{th:LaSalle}.
\end{lem}
\begin{IEEEproof}
The variation of $V(\rho)$ along forward trajectories of the system \eqref{eq:DiscreteTimeSystem} is
\beqa
\Delta V(\rho) &=& \Tr\left(\Pi_R \mathcal{T}[\rho]\right) - \Tr(\Pi_R \rho) \nonumber\\
&=& \Tr\left[\Pi_R \left(\sum_k \M \rho \Ma - \rho\right)\right] \label{eq:DeltaV}
\eeqa
Notice that $\Tr(\sum_k \M \rho \Ma - \rho) = 0$, and that $V(\rho)=0$ for all $\rho$'s that have support in $\mathcal{H}_S$. 
%
Let us express $\sum_k \M \rho \Ma - \rho$ in its block form, using the fact that $M_Q=0$ by assuming invariance of $J_S(\mathcal H_I)$. We get
\begin{multline}
\sum_k  M_k \rho M_k^\dagger - \rho = \\
=	\sum_k \bm{\Mb{S}}{\Mb{P}}{0}{\Mb{R}} \bm{\rho_S}{\rho_P}{\rho_P^\dagger}{\rho_R}
	\bm{\Mab{S}}{0}{\Mab{P}}{\Mab{R}} - \rho \\
= \sum_k \left[ \begin{smallmatrix} \Mb{S}\rho_S\Mab{S} + \Mb{P}\rho_P^\dagger \Mab{S} + \Mb{S}\rho_P\Mab{P} + \Mb{P}\rho_R\Mab{P} \\ \Mb{R}\rho_P\Mab{S} + \Mb{R}\rho_R\Mab{P}  \end{smallmatrix}\right. \\
\left.\begin{smallmatrix} \Mb{S}\rho_P\Mab{R} + \Mb{P}\rho_R\Mab{R} \\  \Mb{R}\rho_R\Mab{R} \end{smallmatrix}\right] 
- \left[\begin{smallmatrix}{\rho_S}&{\rho_P}\\{\rho_P^\dagger}&{\rho_R}\end{smallmatrix}\right] \label{eq:BlockForm}
\end{multline}
Therefore 
\beqa
\Delta V(\rho) &=& \Tr\left[\Pi_R \left(\sum_k M_k \rho M_k^\dagger - \rho\right)\right]\nonumber\\ &=& \Tr\left[\sum_k M_{k,R}\rho_RM_{k,R}^\dagger - \rho_R\right]\,, \label{eq:expressionDeltaV}
\eeqa
so that in order to get $\Delta V \le 0$ the map $\mathcal{T}_R[\rho_R] := \sum_k M_{k,R}\rho_RM_{k,R}^\dagger$ has to be trace non-increasing.

Note that this condition is automatically verified, once $\mathcal T$ is a TPCP map. Indeed, consider the application of $\mathcal T$ on a state $\bar\rho$ which has support on $\mathcal H_R$. According to the block from in \eqref{eq:BlockForm} we have that the total trace of $\mathcal T[\bar\rho]$ is
$$
\Tr\left(\mathcal T[\bar \rho]\right) = \Tr\left(\sum_k \Mb{P}\bar\rho_R\Mab{P}\right)
+ \Tr\left( \sum_k \Mb{R}\bar\rho_R\Mab{R}\right).
$$
Therefore, as both the terms are positive, being $\bar\rho_R\ge 0$, and as $\mathcal T$ is TP, we have for any $\bar \rho_R \in \mathcal D(\mathcal H_R)$
$$
\Tr\left( \sum_k \Mb{R}\bar\rho_R\Mab{R}\right) \le \Tr\left(\mathcal T[\bar \rho]\right) = \Tr\left(\bar\rho_R\right)
$$
and thus $\mathcal T_R$ is trace non-increasing.
\end{IEEEproof}


This leaves us with determining when $\mathcal J_S(\mathcal H_I)$ contains the largest invariant set in $E$. We shall derive conditions that ensure that no other invariant set $W$ exists in $E=\{\rho | \Delta V(\rho) = 0\}$ such that $\mathcal J_S(\mathcal H_I) \subset W$.
We start by giving some preliminary results.

\begin{lem} \label{stationary}
	Let  $\mathcal T$ be a TPCP transformation described by the Kraus map \eqref{eq:KrausMap}. Consider 
		an orthogonal subspace decomposition ${\cal H}_S\oplus{\cal H}_R.$
	Then the set $\mathcal J_R(\mathcal H_I)$ contains an invariant subset if and only if it contains an invariant state.
\end{lem}
\begin{IEEEproof}
The ``if'' part is trivial. On the other hand, $\mathcal J_R(\mathcal H_I)$ is convex and compact, hence if it contains an invariant subset $W$ it also contains the closure of its convex hull, call it $\bar W$. The map $\cal T$ is linear and continuous, so the convex hull of an invariant subset is invariant, and so is its closure. Hence, by Brouwer's fixed point theorem \cite{ziedler-functional} it admits a fixed point $\bar\rho\in \bar W \subseteq \mathcal J_R(\mathcal H_I).$
\end{IEEEproof}

\begin{lem} \label{JWisInvariant}
Let $W$ be an invariant subset of $\mathcal D(\mathcal H_I)$ for the TPCP transformation $\mathcal T$, and let
$$
\mathcal H_W = \supp(W) = \bigcup_{\rho \in W} \supp(\rho).
$$
Then $\mathcal J_W(\mathcal H_I)$ is invariant.
\end{lem}
\begin{IEEEproof}
The proof follows the one for the continuous-time case in \cite[Lemma 8]{ticozzi-markovian}. 
Let $\hat W$ be the convex hull of $W$. By linearity of dynamics it is easy to show that $\hat W$ is invariant too. Furthermore, from the definition of $\hat W$, there exists a $\hat \rho \in \hat W$ such that $\supp(\hat \rho) = \supp(\hat W) = \mathcal H_W$. Consider the decomposition $\mathcal H_I = \mathcal H_W \oplus \mathcal H_W^\perp$, and the corresponding matrix partitioning 
$$
X = \begin{bmatrix}
X_W & X_L \\
X_M & X_N
\end{bmatrix}.
$$
With respect to this partition, $\hat \rho_W$ is full rank while $\hat \rho_{L,M,N}$ are zero blocks. The state $\hat \rho$ is then mapped by $\mathcal T$ according to \eqref{eq:evolutionForInvariance} and therefore, as $\hat \rho_W$ is full rank, it has to be $M_{k,Q} = 0$ for all $k$'s. Comparing it with the conditions given in proposition \ref{pro:Invariance}, we then infer that $\mathcal J_W(\mathcal H_I)$ is invariant.
\end{IEEEproof}

\begin{pro} \label{pr:HRasakernel}
	Consider $\rho\in\mathcal{J}_R(\Hi_I)$ and evolving under the TPCP transformation $\mathcal T$ described by the Kraus map \eqref{eq:KrausMap}. Let the matrices $\M$ be expressed in the block form
	$$
	\M = \bm{\Mb{S}}{\Mb{P}}{0}{\Mb{R}}
	$$
	according to the state space decomposition ${\cal H}_S\oplus{\cal H}_R$, with $\mathcal J_S(\mathcal H_I)$ invariant.
	Then $\rho \in E = \{\rho \in \mathcal D(\mathcal H_I) | \Delta V(\rho) = 0\}$, where $V(\rho)$ is defined by \eqref{eq:LyapFunction},
	if and only if its $\rho_R$ block satisfies
	$$
	\text{supp}(\rho_R) \subseteq \bigcap_k \ker \left( \Mb{P} \right).
	$$
\end{pro}
\begin{IEEEproof}
	By direct computation, see \eqref{eq:BlockForm}, we have
	\begin{equation} \label{eq:evolutionOfRhoinHR}
	\sum_k  M_k \rho M_k^\dagger = 
		\sum_k \begin{bmatrix} \Mb{P}\rho_R\Mab{P} & \Mb{P}\rho_R\Mab{R} \\
		\Mb{R}\rho_R\Mab{P} & \Mb{R}\rho_R\Mab{R} \end{bmatrix} 
	\end{equation}
	as $\rho$ has support on $\mathcal H_R$ alone.
	Note that as $V(\rho) = 1$, $\Delta V(\rho)=0$ is equivalent to $\mathcal T[\rho]$ having support on $\mathcal H_R$.
	Given the form of the upper-left block of \eqref{eq:evolutionOfRhoinHR}, this is true if and only if $\text{supp}(\rho_R) \subseteq \bigcap_k \ker \left( \Mb{P} \right)$.
\end{IEEEproof}

Proposition \ref{pr:HRasakernel} allows then to state the following key characterization of global, asymptotical stability of $\mathcal {J}_S(\Hi_I)$.

\begin{thm} \label{th:attractivity}
	Let the TPCP transformation $\mathcal T$ be described by the Kraus map \eqref{eq:KrausMap}. Consider an orthogonal subset decomposition ${\cal H}_S\oplus{\cal H}_R$, with $\mathcal J_S(\mathcal H_I)$ invariant. Let the matrices $\M$ be expressed in their block form
	$$
	\M = \bm{\Mb{S}}{\Mb{P}}{0}{\Mb{R}}
	$$
	according to the same state space decomposition.
	Then the set $\mathcal J_S(\mathcal H_I)$ is GAS if and only if there are no invariant states with support on 
	\begin{equation*}
		\bigcap_k \ker \left( \Mb{P} \right).
	\end{equation*}
\end{thm}
\begin{IEEEproof}
Necessity is immediate: if there was an invariant state with support on $\bigcap_k \ker \left( \Mb{P} \right)$, It would have non trivial support on $\mathcal H_R$, and therefore $\mathcal H_S$ could not be attractive. In order to prove the other implication, consider LaSalle's theorem. By hypotesis, $\mathcal J_S(\mathcal H_I)$ is invariant and is contained in $E$ (see Proposition \ref{pr:HRasakernel}), therefore it is contained in the largest invariant set $W$ in the zero-difference locus $E$.

Let us suppose that $\mathcal J_S(\mathcal H_I) \subset W$, but $\mathcal J_S(\mathcal H_I) \ne W$. That is, there exists a set $W \subseteq E$ which is invariant and strictly contains $\mathcal J_S(\mathcal H_I)$. Therefore its support has to be 
$$
\mathcal H_W = \mathcal H_S \oplus \mathcal H_{R'}
$$
with $\mathcal H_{R'}$ subspace of $\mathcal H_{R}$, and by Lemma \ref{JWisInvariant} $\mathcal J_W(\mathcal H_I)$ must be invariant too. Consider then a state $\hat \rho$ which belongs to $\mathcal J_W(\mathcal H_I)$, with non trivial support on $\mathcal H_{R'}$, and define
$$
\tilde \rho = \frac{\Pi_{R'}\hat \rho \Pi_{R'}}{\Tr(\Pi_{R'}\hat \rho)} = \bm{0}{0}{0}{\tilde \rho_R}
$$ 
which has support on $\mathcal H_{R'}$ only. By construction, $\tilde \rho$ is in $\mathcal J_W(\mathcal H_I)$, and therefore its trajectory is contained in $\mathcal J_W(\mathcal H_I)$. It is also in $E$, that is $\Delta V (\tilde \rho)=0$. As we have $V(\tilde \rho) = 1$, then its evolution must be also remain in $\mathcal H_{R'}\subseteq \Hi_R$ at any time. Therefore an invariant set with support on $\mathcal H_{R}$ exists.
By reversing the implication, this means that if does not exist an invariant set with support on $\mathcal H_{R}$, then  $\mathcal J_S(\mathcal H_I)$ is the largest invariant set in $E$. Furthermore, Proposition \ref{pr:HRasakernel} indicates that if there is an invariant set in $E$ with support on $\mathcal H_{R},$ its support must actually be contained in $$\bigcap_k \ker \left( \Mb{P} \right).$$
Therefore, if no such subset exists, we have attractivity of $\mathcal J_S(\mathcal H_I)$ by LaSalle's theorem.  By Lemma \ref{stationary}, the existence of an invariant set is equivalent to the existence of an invariant state with support on $\bigcap_k \ker \left( \Mb{P} \right)$.
\end{IEEEproof}

Given the usual decomposition ${\cal H}_I={\cal H}_S\oplus{\cal H}_R,$ let us further decompose $\mathcal H_R$ in
$$
{\cal H}_{R'}= {\cal H}_{R}\ominus\bigcap_k \ker \left( \Mb{P} \right)\quad \text{and} \quad{\cal H}_{R''}=\bigcap_k \ker \left( \Mb{P} \right) 
$$
and consider the operation elements $\M$ in a basis induced by the decomposition ${\cal H}_I={\cal H}_S\oplus{\cal H}_{R'}\oplus{\cal H}_{R''},$:
$$
	\M =
	\left[\begin{array}{c|c|c}
		\Mb{S}	&	\Mb{P'} & 0 \\
		\hline
		0		&	\Mb{R1} & \Mb{R2}\\
		\hline
		0		&	\Mb{R3} & \Mb{R4}
	\end{array}\right]\,.
$$
Density operators $\rho$ which have support on the bottom right block clearly belong to $\mathcal J_{R''}(\mathcal H_I)$. Sufficient, although not necessary, condition to be sure that no invariant sets have support on that subspace is that
$$
\bigcap_k \ker \left( \Mb{R2} \right) = \{0\}\quad \text{ and } \quad  \Mb{R3} = 0 \; \forall k\,.
$$
This way, the states that have support on $\bigcap_k \ker \left( \Mb{P} \right)$ will be mapped into states which has non-trivial support on $\left[ \bigcap_k \ker \left( \Mb{P} \right) \right]^\perp$, and therefore no invariant set will exist in $\bigcap_k \ker \left( \Mb{P} \right)$. 
This intuition will be further developed in the Section \ref{engineering}, where a control design tool capable of achieving attractivity of a given subspace is obtained. 


\section{A canonical matrix form based on the QR decomposition}
\label{sec:canonicalForm}

\subsection{On the uniqueness of QR decomposition}

\begin{dfn}[QR decomposition \cite{horn-johnson}]
A QR decomposition of a complex-valued square matrix $A$ is a decomposition of $A$ as
$$
   A = QR,
$$
where $Q$ is an orthogonal matrix (meaning that $Q^\dagger Q = I$ ) and $R$ is an upper triangular matrix.
\end{dfn}

We investigate here the uniqueness of the QR decomposition of a complex-valued matrix $A$, both in the case in which $A$ is non-singular and in the case in which it is singular. While the real-matrix case is well known (see e.g. \cite{horn-johnson}), a little extra care is needed in the complex case. Let's first analyze the case of $A$ non-singular. We have the following result.

\begin{lem}\label{lemm:uniquenessOfQR_nonsingular}
Let
$$
A = Q_1R_1 \,,\quad A = Q_2R_2
$$
be two QR decompositions of the same \emph{non-singular} square matrix $A$. Then $R_1$ and $R_2$ only differ for the phase of their rows, that is
$$
R_1 = \Phi R_2\,, \qquad Q_1 = Q_2 \Phi^{-1}
$$
where $\Phi = \text{diag}\left(e^{j\phi_1}, \ldots, e^{j\phi_n}\right)$.
\end{lem}
\begin{IEEEproof}
Using the fact that $A$ and its factors are non singular, we have
$$
Q_2^\dagger Q_1 = R_2 R_1^{-1} = \Phi
$$
where $\Phi$ is upper triangular because it is the product of two upper triangular matrices. The matrix $\Phi$ must also be orthonormal, because it is the product of two orthonormal matrices. Therefore, starting from the first column of $\Phi$, we have that
$
\lvert \Phi_{11} \rvert = 1.
$
where $\Phi_{ij}$ is the element of $\Phi$ in position $(i,j)$.

We proceed by induction on the column index $j$. Assume that all the columns $\Phi_k$ with $k<j$ satisfy $\Phi_{lk}=0$ for any $l \ne k$. In order for $\Phi_j$ to satisfy $\Phi_k^\dagger \Phi_j = 0\; \forall k<j$, it must be
$$
\Phi_{1j} = \cdots = \Phi_{j-1,j} = 0.
$$
Moreover, as $\Phi$ is upper triangular, we must also have
$
\Phi_{j+1,j} = \cdots = \Phi_{n,j} = 0.
$
Therefore, by othonormality of $\Phi$, it has to be
$
\lvert \Phi_{jj} \rvert = 1.
$
\end{IEEEproof}

In the case in which $A$ is singular, on the other hand, the QR decomposition is not just unique up to a phase of the rows of $R$.

{\em Example 1.} Consider the following matrix:
$$M=\left[\begin{array}{cc}0 & 1  \\0 & 1 \end{array}\right].$$
Since it is already upper triangular, a valid QR decomposition is given by $Q=I,\,R=M.$
On the other hand, we can consider
$$Q=\frac{1}{\sqrt{2}}\left[\begin{array}{cc}1 & 1  \\1 & -1 \end{array}\right],\; R=\left[\begin{array}{cc}0 & \sqrt{2}  \\0 & 0 \end{array}\right],$$ which clearly also give $QR=M,$ with $Q^\dag Q=I.$

However, introducing some conditions on the QR decomposition, it is possible to obtain a \emph{canonical form} for the QR decomposition in a sense that will be explained later in this section. A useful lemma in this sense is the following.

\begin{lem} \label{lemm:subspacesInCanonicalForm}
Consider a QR decomposition of a square matrix $A$ of dimension $n$, and an index $\bar j$ in $[1,n]$, such that
\begin{equation}\label{CanonicalFormCondition}
r_{ij} = 0 \quad \forall j \le \bar j, \forall i>\rho_j
\end{equation}
where $\rho_j$ is the rank of the first $j$ columns of $A$.
Let $a_i$ and $q_i$, be the $i$-th column of $A$ and $Q$ respectively. Then
$$
<a_1, \ldots, a_j> \,=\, <q_1, \ldots, q_{\rho_j}> \quad \forall j=1,\ldots,\bar j.
$$
\end{lem}

\begin{IEEEproof}
Consider the expression for the $j$-th column of $A$
$$
a_j = Q r_j.
$$
By the hypothesis, the last $n-\rho_j$ elements of $r_j$ are zeros, hence it results
$$
a_j\,\in\, <q_1, \ldots, q_{\rho_j}> \quad \forall j=1,\ldots,\bar j
$$
and therefore
$$
<a_1, \ldots, a_j> \,\subseteq\, <q_1, \ldots, q_{\rho_j}> \quad \forall j=1,\ldots,\bar j.
$$
As the rank of the first $j$ columns is $\rho_j$, which is also the dimension of $<q_1, \ldots, q_{\rho_j}>$, equality of the two subspaces holds.
\end{IEEEproof}

We next show as a particular choice of the QR decomposition, suggested by lemma \ref{lemm:subspacesInCanonicalForm}, gives a canonical form on $\C^{n\times n}$ with respect to left-multiplication for elements of the unitary matrix group $\mathcal{U}(n)$.
We construct the QR decomposition through the Gram-Schmidt orthonormalization process, fixing the degrees of freedom of the upper-triangular factor $R$ and verifying that the resulting decomposition satifies the hypotesis of lemma \ref{lemm:subspacesInCanonicalForm} for $\bar j = n$.

\subsection{Construction of the QR decomposition by orthonormalization}

\begin{thm} \label{th:QRconstruction}
Given any (complex) square matrix $A$ of dimension $n,$ it is possible to derive a QR decomposition $A= QR$ such that hypoteses of Lemma \ref{lemm:subspacesInCanonicalForm} are satisfied, and such that the first nonzero element of each row of $R$ is real and positive.
\end{thm}

\begin{IEEEproof} We explicitly construct the QR decomposition of $A$ column by column.  We denote by $A,Q,R$ the matrices, with $a_i,q_i,r_i$ their $i$-th columns and with $a_{i,j},q_{i,j},r_{i,j}$ their elements, respectively. Let us start from the first non zero column of $A\in\mathbb{C}^{n\times n}$, $a_{i_0}$, and define
\begin{equation}\label{definitionQ1}
	q_1 = \frac{a_{i_0}}{\|a_{i_0}\|}, \qquad r_{1,{i_0}} = \|a_{i_0}\|, \qquad r_{2,i_0} = \ldots = r_{n,i_0}=0.
\end{equation}
Also fix $r_{j}=0$ for all $j < {i_0}$. 

The next columns of $Q,R$ are constructed by an iterative procedure. Define $\rho_{i-1}$ as the rank of the first ${i-1}$ columns of $A.$ We can assume (by induction) to have the first $\rho_{i-1}$ columns of $Q$ and the first ${i-1}$ columns of $R$ constructed in such a way that $r_{k,j}=0$ for $k>\rho_{j}$ and $j \le {i-1}$.

Consider the next column of $A,$ $a_{i}$. Assume that $a_{i}$ is linearly dependent with the previous  columns of $A$, that is $\rho_{i} = \rho_{i-1}$. Since lemma \ref{lemm:subspacesInCanonicalForm} applies, $a_{i}$ can be written as 
$$
a_{i} = \sum_{j=1}^{i-1} \alpha_j a_j = \sum_{j=1}^{i-1} \alpha_j \sum_{\ell=1}^{\rho_j} r_{\ell,j} q_\ell
$$
and therefore, being $a_{i}$ a linear combination of the columns $\{q_1,\ldots, q_{\rho_{i-1}}\}$, the elements of $r_{i}$ are defined as
$$
r_{\ell,i} = q_\ell^\dagger a_{i}\,,\quad \text{ for }\ell=1,\ldots,\rho_{i}.
$$

On the other hand, if the column $a_{i}$ is linearly independent from the previous columns of $A$, then the rank $\rho_{i} = \rho_{i-1}+1$. As before, the first $\rho_{i-1}$ coefficients of $r_{i}$ must be defined as
$$
r_{\ell,i} = q_\ell^\dagger a_{i}\,,\quad \text{ for }\ell=1,\ldots,\rho_i -1,
$$
and let set $r_{\ell,i} = 0\, \text{ for }\ell=\rho_{i}+1,\ldots,n.$
Let us also introduce 
\begin{equation} \label{definitionTildeA}
\tilde a_{i} := a_{i} - \sum_{\ell=1}^{\rho_i} r_{\ell,i} q_\ell \ne 0
\end{equation}
and define
\begin{equation} \label{definitionQRHOi}
q_{\rho_{i}} = \frac{\tilde a_{i}}{\|\tilde a_{i}\|}\, \quad r_{\rho_{i},i} = \|\tilde a_{i}\|,
\end{equation}
It is immediate to verify that the obtained $q_{\rho_i}$ is orthonormal to the columns $q_1,\ldots,q_{\rho_i - 1}$, and that $a_i = Q r_{\rho_i}$.

After iterating until the last column of $R$ is defined, we are left to choose the remaining columns of $Q$ so that the set $\{q_1,\ldots,q_n\}$ is an orthonormal basis for $\mathbb{C}^{n\times n}$. By construction, $A=QR$.
\end{IEEEproof}

\subsection{$R$ is a canonical form}

Let $\mathcal{G}$ be a group acting on $\C^{n\times n}.$ Let $A,B\in \C^{n\times n}.$ If there exists a $g\in{\cal G}$ such that $g(A)=B,$ we say that $A$ and $B$ are ${\cal G}$-equivalent, and we write $A\sim _{\cal G} B.$ 
\begin{dfn}\label{canonicalform} A canonical form  with respect to $\mathcal{G}$ is a function $\mathcal{F}:\C^{n\times n}\rightarrow \C^{n\times n}$ such that for every $A,B\in\C^{n\times n}$:\begin{itemize}
\item[i.]${\cal F}(A)\sim _{\cal G} A $;
\item[ii.] ${\cal F}(A)={\cal F}(B)$ if and only if $A\sim _{\cal G} B.$
\end{itemize}
\end{dfn}

Let us consider the unitary matrix group $\mathcal{U}(n)\subset\C^{n\times n}$ and consider its action on $\C^{n\times n}$ through left-multiplication, that is, for any $U\in\mathcal{U}(n),\,M\in\C^{n\times n}$:
$$U(M)=UM.$$

We are now ready to prove the following.

\begin{thm} \label{th:Canonical}
	Define $\mathcal{F}(A)=R$, with  $R$ the upper-triangular matrices obtained by the procedure described in the proof of theorem \ref{th:QRconstruction}. Then $\mathcal{F}$ is a canonical form with respect to ${\cal U}(n)$ (and its action on $\mathbb{C}^{n\times n}$ by left multiplication).
\end{thm}
\begin{IEEEproof}
By construction $A=QR,$ with unitary $Q,$ so $\mathcal{F}(A)\sim_{{\cal U}(n)}A.$
If $A,B\in \mathcal{C}^{n\times n}$ are such that $\mathcal{F}(A)=\mathcal{F}(B)=R,$ thus $A=QR$ and $B=VR$ for some $Q,V\in{\cal U}(n),$ and hence $A=QV^{-1}B.$ 

On the other hand,
if $A=UB,$ $U\in{\cal U}(n),$ we have to prove that the upper-triangular matrix in the canonical QR decompositions $A=QR^{(A)}$ and $B=VR^{(B)}$ is the same. 
If the first non zero column of $B$ is $b_{i_0},$ then the first column nonzero column of $A$ is, being $U$ unitary, $a_{i_0}=U b_{i_0}.$ One then finds from \eqref{definitionQ1}
\begin{equation}
v_1 = \frac{U^\dag a_{i_0}}{ \|U^\dag a_{i_0}\|}=U^\dag q_{1}  \quad
r^{(B)}_{1,{i_0}} = \|U^\dag a_{i_0}\| = r^{(A)}_{1,{i_0}}.
\end{equation}
Hence the first ${i_0}$ columns of $R^{(A)}$ and $R^{(B)}$ are identical.
We then proceed by induction. 
Assume that $r^{(A)}_{j} = r^{(B)}_{j},\;q_{j} = Uv_{j}$ for $j=1,...,i-1$.
If the column $a_i$ is linearly dependent from the previous $i-1$ so it must be $b_i.$ The elements of $r^{(A)}_{i}$ are defined as
$$
r^{(A)}_{k,i} = q_k^\dagger a_i=q_k^\dagger U U^\dag a_i = v_k^\dag b_i = r^{(B)}_{k,i}\,,\quad \text{ for }k=1,\ldots,\rho_i-1.
$$
On the other hand, if the column $a_i$ is linearly independent from the previous columns of $A$, then the rank $\rho_i = \rho_{i-1}+1$. As before, the first $\rho_i -1$ coefficients of $r_i$ are defined as
$$
r^{(A)}_{k,i} = q_k^\dagger a_i=q_k^\dagger U U^\dag a_i = v_k^\dag b_i = r^{(B)}_{k,i}\,,\quad \text{ for }k=1,\ldots,\rho_i -1,
$$
and $r^{(A)}_{k,i} = r^{(B)}_{k,i} = 0\, \text{ for }k=\rho_i+1,\ldots,n$.
Let us consider as before
$$
\tilde a_i := a_i - \sum_{k=1}^{\rho_i -1} r^{(A)}_{ki} q_k \ne 0.
$$
By using the equivalent definition and the inductive hypothesis it follows that $\tilde b_i=U^\dag \tilde a_i$ and 
$$
v_{\rho_i} = \frac{U^\dag \tilde a_i}{\|U^\dag \tilde a_i\|} = U^\dag q_{\rho_i}   \quad 
r^{(B)}_{\rho_i,i} = \|U^\dag\tilde a_i\|=r^{(A)}_{\rho_i,i}.
$$
Hence $r^{(A)}_{i}=r^{(B)}_{i},$ and by induction $R^{(A)}=R^{(B)}.$

\end{IEEEproof}


\section{Engineering attractive subspaces via closed-loop control}\label{engineering}
\subsection{The controlled dynamics}

In this section we deal with the problem of stabilization of a given quantum subspace by discrete-time measurements and unitary control. The control scheme we employ follows the ideas of \cite{viola-engineering, ticozzi-feedbackDD}. 

Suppose that a generalized measurement operation can be performed on the system at times $t=1,2,\ldots$, resulting in an open system, discrete-time dynamics described by a given Kraus map, with associated Kraus operators $\{M_k\}$. This can be realized, for example, when the system is coupled to an auxiliary measurement apparatus, it is manipulated coherently, and then a projective measurement is performed on the auxiliary system (see appendix \ref{app:QuantumMeasurements}). Suppose moreover that we are allowed to unitarily control the state of the system, i.e. $\rho_{controlled}=U\rho U^\dag,$ $U\in{\mathcal U}(\mathcal H_I)$.  We shall assume that the control is fast with respect to the measurement time scale, or the measurement and the control acts in distinct time slots. 

\begin{figure}[htbp]
\centering
\includegraphics[width=0.5\textwidth]{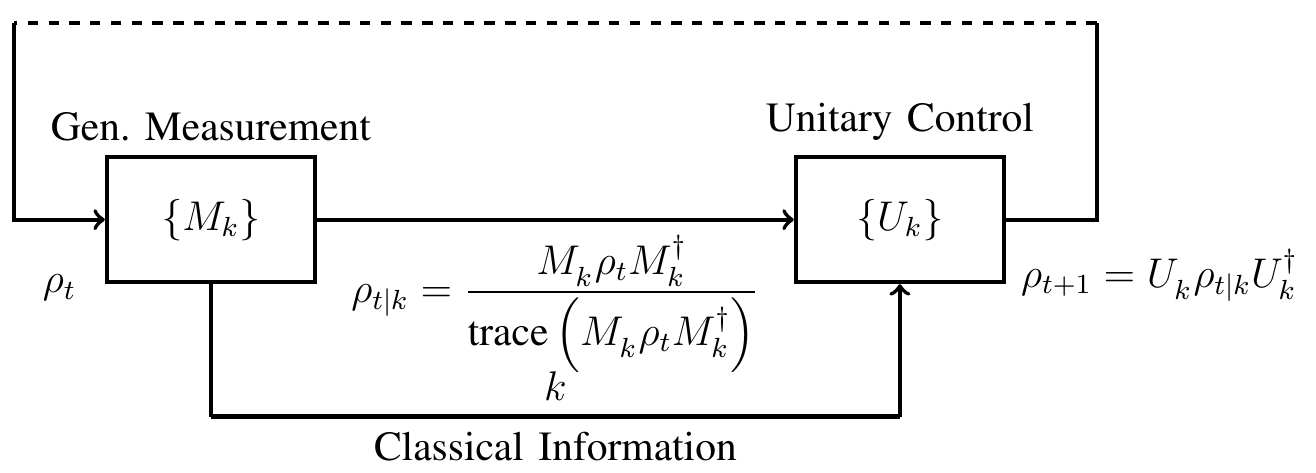} 
\caption{A measurement-dependent unitary control scheme.}
\label{figsel} 
\end{figure}

We can then use the generalized measurement outcome $k$ to condition the control choice, that is, a certain coherent transformation $U_k$ is applied after the $k$-th output is recorded. The measurement-control loop is then iterated: If we average over the measurement results at each step, this yields a different TPCP map, which dependens on the design of the set of unitary controls $\{U_k\}$ and describes the evolution of the state {\em immediately after} each application of the controls:
$$ \rho(t+1)=\sum_kU_kM_k\rho(t)M_k^\dag U_k^\dag. $$
Figure \ref{figsel} depicts the feedback control loop (before the averaging).

%

In this section we will tackle the problem of characterizing the set of open loop dynamics that can be engineered through this feedback setup {\em by designing the set $\{U_k\}$ with fixed measurement operator $\{M_k\}$}. Moreover, on the basis of the analysis results of section \ref{gas}, we will derive an algorithm that allows to design the set of unitary controls such that global asymptotical stability of a {\em given} subspace is achieved. Notice that the desired result is going to be achieved for the averaged time-evolution that describes the system immediately after the control step. It is easy to show that if a certain subspace is GAS  for the averaged dynamics, it must be so also for the conditional ones.

\subsection{Simulating generalized measurements}

A first straightforward application of the canonical form we derived in the previous section is the following. Assume we are able to perform a generalized measurement, with associated operators $\{M_k\}_{k=1}^m,$ and we would like to actually implement a different measurement with associated operators $\{N_k\}_{k=1}^m,$ by using the unitary control loop as above. Notice that the control scheme we considered allows to modify only the conditioned states, not the probability of the outcomes, since
$ \trace(M_k^\dag M_k \rho)=\trace (M_k^\dag U_k^\dag U_k M_k \rho).$
The following holds:

\begin{pro}
A measurement with associated operators $\{N_k\}_{k=1}^m$ can be simulated by a certain choice of unitary controls from a measurement $\{M_k\}_{k=1}^m,$ if and only if there exist a reordering $j(k)$ of the first $m$ integers such that:
$$
	{\mathcal F}(N_k)={\mathcal F}(M_{j(k)}),
$$
where ${\mathcal F}$ returns the canonical $R$ factor of the argument, as described in the section \ref{sec:canonicalForm}.
\end{pro}
\begin{IEEEproof}
Let us first assume that for a given reordering $j(k)$ it holds ${\mathcal F}(N_k)={\mathcal F}(M_{j(k)})=R_k$. Therefore the canonical QR decomposition of $N_k$ and $M_{j(k)}$ gives
$$
N_k = U_k R_k\,,
\qquad
M_{j(k)} = V_{j(k)} R_{k}.
$$
Let then $U_k V_{j(k)}^\dag$ be the unitary control associated with the measurement outcome $k$. We have
\beqan
\mathcal T_\text{closed loop}[\rho]
&=& \sum_k U_k V_{j(k)}^\dag M_{j(k)} \rho M_{j(k)}^\dag V_{j(k)} U_k^\dag 
\\%
&=& \sum_k U_k R_k \rho R_k^\dag U_k^\dag 
= \sum_k N_k \rho N_k^\dag 
\eeqan
and therefore it simulates the measurement associated with the operators $\{N_k\}_{k=1}^m$.

On the other hand, suppose that there exists a set of unitary controls $\{Q_k\}_{k=1}^m$ and there is a reordering $j(k)$ of the first $m$ integers such that
$$
Q_{j(k)} M_{j(k)} = N_k.
$$
According to Theorem \ref{th:Canonical}, $\mathcal{F}$ is a canonical form with respect to ${\cal U}(n)$ and its action on $\mathbb{C}^{n\times n}$ by left multiplication, and therefore if $Q_{j(k)} M_{j(k)} = N_k$ then ${\cal F}(M_{j(k)}) = {\cal F}(N_k).$
\end{IEEEproof}

\subsection{Global asymptotic stabilization of a quantum subspace}

Suppose that the operators $\{\M\}$ are given, corresponding to a measurement that is performed on the quantum system, with corresponding outcomes $\{k\}$. We are then looking for a set of unitary transformations $\{\mat{U}\}$ such that, once they are applied to the system, the resulting transformation
$$
\mathcal T[\rho] = \sum_k \mat{U}\M\rho\Ma\mata{U}
$$
makes a given subsystem $\mathcal S$ GAS. 
Let us introduce a preliminary result.

\begin{lem}\label{unitaryThatGuaranteesInvariance}
	Let $R$ be the upper triangular factor of a \emph{canonical QR decomposition} in the form
	$$
	R = \bm{R_S}{R_P}{0}{R_R}
	$$
	(according to the block structure induced by \eqref{eq:spaceDecomposition}) and suppose $R_P = 0$. Consider the matrix $N$ obtained my left multiplying $R$ by a unitary matrix $V$:
	$$
	N = VR = \bm{V_S}{V_P}{V_Q}{V_R}\bm{R_S}{0}{0}{R_R} = \bm{N_S}{N_P}{N_Q}{N_R}.
	$$
	Then $N_Q = 0$ implies $N_P = 0.$
\end{lem}
\begin{IEEEproof}
	Let first consider the case in which $R_S$ has full rank.
	Let $r \times m$ be the dimension of $V_Q$, and $m \times m$ be the dimension of $R_S$.
	Since it must be $N_Q = V_Q R_S = 0$  and $R_S$ is full rank, we have
	$
	 V_Q = 0.
	$

   As $V$ is unitary, its column must be orthonormal.
 Being $V_Q = 0$, $V_S$ must be itself an orthonormal block in order to have orthonormality of the first $m$ columns of $V$.
	It then follows that $V_P=0$, because any $j$-th column, $j > m$, must be orthonormal to all the first $m$ columns.
	It then follows that
	$$
	N_P = V_P R_R = 0.
	$$
	
	Let us now consider the other case, in which $R_S$ is singular. This implies that $\rho_m < m$ ($\rho_m$ being the rank of the first $m$ columns of $R$): Therefore, as $R$ is a triangular factor of a canonical QR decomposition, the element $R_{\rho_m+1,m}=0.$
	
	Now, by construction of the canonical QR decomposition, if there were non-zero columns of index $j>m,$ one of them would have a non-zero element on the row of index $\rho_m + 1$. By recalling that $R_P = 0,$ we have that $R_{\rho_m + 1,j} = 0,$ $\forall j\in[m+1,m+r]$.
	Therefore, all the last $r$ columns are zero-vectors, and in particular $R_R = 0$. It then follows that
	$$
	N_P = V_P R_R = 0.
	$$
\end{IEEEproof}



%

This result will be instrumental in proving the main theorem of the section, which provides us with an iterative control design procedure that renders the desired subspace asymptotically stable whenever it is possible.

\begin{thm}\label{th:algorithm}
Consider a subspace orthogonal decomposition $\Hi_I=\Hi_S\oplus\Hi_R$ and a given generalized measurement associated to Kraus operators $\{M_k\}.$ If asymptotic stability of a subspace $\mathcal S$ can be achieved by any measurement-dependent unitary control $\{U_k\}$, it can be achieved by building $U_k$ using the iterative algorithm below.
\end{thm}

\begin{quotation}
\noindent \rule{\linewidth}{1pt} \\
{\em Control design algorithm} \\[-2mm]
\noindent \rule{\linewidth}{1pt} 

Let $\{|\phi\rangle_j^S\}_{j=1}^m$, $\{|\phi\rangle_k^R\}_{k=1}^r$ denote orthonormal bases for $\mathcal H_S$ and $\mathcal H_R$, and represent each $M_k$ as a matrix with respect to the basis $\{|\phi\rangle_j^S\}_{j=1}^m\bigcup\{|\phi\rangle_k^R\}_{k=1}^r.$
Compute a QR decomposition $M_k=Q_kR_k$ with canonical $R_k$ for each $k.$
Call $\Hi_R^{(0)}=\Hi_R,$ $\Hi_S^{(0)}=\Hi_S,$ $U_k^{(0)}=Q_k^\dagger$ and rename the matrix blocks $R_{S,k}^{(0)}=R_{S,k},$ $R_{P,k}^{(0)}=R_{P,k}$ and $R_{R,k}^{(0)}=R_{R,k}.$ 

If $R_{P,k} = 0 \; \forall k$, then the problem is not feasible and a unitary control law cannot be found. Otherwise define $V^{(0)}=I$, $Z^{(0)}=I$, and consider the following iterative procedure, starting from $i=0$:

\begin{enumerate}
\item Define $ \Hi_R^{(i+1)}=\bigcap_k\ker R_{P,k}^{(i)}:$\\
If $ \Hi_R^{(i+1)}=\{0\}$ then the iteration is successfully completed. Go to step 8).\\
If $ \Hi_R^{(i+1)}\subsetneq \Hi_R^{(i)},$ define $\Hi_S^{(i+1)}=\Hi_R^{(i)}\ominus\Hi_R^{(i+1)}$ and $Y^{(i+1)}=I$.\\
If $ \Hi_R^{(i+1)}= \Hi_R^{(i)}$ (i.e. $R_{P,k}^{(i)} = 0 \; \forall k$) then, if $\dim (\Hi_R^{(i)}) \geq \dim (\Hi_S^{(i)})$:

\begin{enumerate} 

\item Choose a subspace $\Hi_S^{(i+1)}\subseteq \Hi_R^{(i)}$ of the same dimension of $\Hi_S^{(i)}.$ (Re)-define $\Hi_R^{(i+1)}=\Hi_R^{(i)}\ominus \Hi_S^{(i+1)}.$

\item Let $\Hi^{(i)}_T=\bigoplus_{j=0}^{i-1}\Hi_S^{(j)}.$ Construct a unitary matrix $Y$ with the following block form, according to a Hilbert space decomposition $\Hi_I=\Hi^{(i)}_T\oplus \Hi_S^{(i)}\oplus \Hi_S^{(i+1)} \oplus \Hi_R^{(i+1)} $:
\begin{equation}
Y^{(i+1)} = 
\left[\begin{array}{c|c|c|c}I & 0 & 0 & 0 \\\hline0 & 1/\sqrt{2}I  & 1/\sqrt{2}I & 0 \\\hline 0 & 1/\sqrt{2}I  & -1/\sqrt{2}I & 0 \\\hline 0 & 0 & 0 & I\end{array}\right].
\label{eq:YwhenNotIdentity}
\end{equation}

\end{enumerate}

If instead $\dim (\Hi_R^{(i)}) < \dim(\Hi_S^{(i)})$:

\begin{enumerate} 

\item Choose a subspace $\Hi_S^{(i+1)}\subseteq \Hi_S^{(i)}$ of the same dimension of $\Hi_R^{(i)}.$

\item Let $\Hi^{(i)}_T=\left(\bigoplus_{j=0}^{i-1}\Hi_S^{(j)}\right) \oplus \left(\Hi_S^{(i)}\ominus \Hi_S^{(i+1)} \right).$ Construct a unitary matrix $Y$ with the following block form, according to a Hilbert space decomposition $\Hi_I=\Hi^{(i)}_T\oplus \Hi_S^{(i+1)}\oplus \oplus \Hi_R^{(i+1)} $:
\begin{equation}
Y^{(i+1)} = 
\left[\begin{array}{c|c|c}I & 0 & 0  \\\hline0 & 1/\sqrt{2}I  & 1/\sqrt{2}I  \\\hline 0 & 1/\sqrt{2}I  & -1/\sqrt{2}I  \end{array}\right].
\label{eq:YwhenNotIdentity1}
\end{equation}

\item Define $Z^{(i+1)}=Z^{(i)}Y^{(i+1)}$ and go to step 8).

\end{enumerate}

\item Define $Z^{(i+1)}=Z^{(i)}Y^{(i+1)}.$

\item Rewrite $\tilde R^{(i)}_{R,k}=W^{(i+1)}R^{(i)}_{R,k}W^{(i+1)\dag}$ in a basis according to the $\Hi_R^{(i)}=\Hi_S^{(i+1)}\oplus\Hi_R^{(i+1)}$ decomposition. 

\item Compute the canonical QR decomposition of $\tilde R_{R,k}^{(i)}=Q_k^{(i+1)}R_{k}^{(i+1)}.$ Compute the matrix blocks $R_{P,k}^{(i+1)},R_{R,k}^{(i+1)}$ of $R_k^{(i+1)}$, again according to the decomposition $\Hi_R^{(i)}=\Hi_S^{(i+1)}\oplus\Hi_R^{(i+1)}$.

\item Define $U^{(i+1)}=\left[\begin{array}{c|c}I & 0 \\\hline 0 &  W^{(i+1) \dag} \left( Q_k^{(i+1)} \right)^\dag W^{(i+1)}\end{array}\right]U^{(i)}.$ 

\item Define $V^{(i+1)}=\left[\begin{array}{c|c}I & 0 \\\hline 0 &  W^{(i+1)} \end{array}\right]V^{(i)}.$

\item Increment the counter and go back to step 1).

\item Return the unitary controls $U_k=V^{(i)\dag}Z^{(i)}V^{(i)}U_k^{(i)}$.

\end{enumerate}

\noindent \rule{\linewidth}{1pt} \\
\end{quotation}

\begin{IEEEproof}
\noindent Let us first consider the case in which the algorithm stops before the iterations. This happens if for every $k$ we have $R_{P,k}=0.$ Remember that each $R_k$ has been put in canonical form, so Lemma \ref{unitaryThatGuaranteesInvariance} applies:  This means that any control choice that ensures invariance of the desired subspace, that is $N_k=U_k R_k$ with $N_{Q,k}=0$, makes also $\mathcal{J}_R(\Hi_I)$ invariant, since $N_{P,k}=0.$ Hence an invariant state with support on $\Hi_R$ always exists. This, via Theorem \ref{th:attractivity}, precludes the existence of a control choice that renders ${\cal J}_S(\Hi_I)$ GAS. 

If the algorithm does not stop, then at each step of the iteration the dimension of $ \Hi_R^{(i)}$ is reduced by at at least $1,$ hence the algorithm is completed in at most $n$ steps. If the algorithm is successfully completed at a certain iteration $j,$ we have built unitary controls $\{U_k^{(j)}\}$ and a unitary $V^{(j)}$ such that the controlled quantum operation element, under the change of basis $V^{(j)},$ is of the form:
\beqa
\tilde N_k &=& 
V^{(j)} U_k M_k V^{(j)\dag} \nonumber\\ &=&
Z^{(j)}
\left[\begin{array}{c|c|c|c|c}
		R_{S,k}^{(0)} & \bar R_{P,k}^{(0)} & 0 & 0 & 0 \\
		\hline 0 & R_{S,k}^{(1)} & \ddots & 0 & 0 \\
		\hline 0 & 0 & \ddots & \bar R_{P,k}^{(j-1)} & 0 \\
		\hline 0 & 0 & 0 & R_{S,k}^{(j)} & \bar R_{P,k}^{(j)} \\
		\hline 0 & 0 & 0 & 0 & R_{R,k}^{(j)}\end{array}\right] 
\label{eq:resultingClosedLoopEvolution}
\eeqa
where the block structure is consistent with the decomposition $\bigoplus_{i=0}^{j+1} \Hi_S^{(i)}$ (where to simplify the notation we set $\Hi^{(j+1)}_{S} =  \Hi^{(j)}_{R}$).
Let $\bar R_k$ be the block matrix above and consider its upper-triangular part. The rows have the form $\begin{bmatrix}\bar R_{P,k}^{(i)} & 0 & \ldots & 0 \end{bmatrix}$ because  at each step of the iteration we choose a basis $W^{(i)}$ according to the decomposition $\Hi_S^{(i+1)}\oplus \Hi_R^{(i+1)},$ where $ \Hi_R^{(i+1)} \subseteq \bigcap_k\ker R_{P,k}^{(i)},$ hence obtaining $R_{P,k}^{(i)}W^{(i)\dag} = \begin{bmatrix}\bar R_{P,k}^{(i)} & 0 & \ldots & 0 \end{bmatrix}$. It is easy to verify that the subsequent unitary transformations have no effects on the blocks $\bar R_{P,k}^{(i)}.$ 

The upper-triangular form of each $\bar R_k$ and the form of $Z^{(j)}$ and $V^{(j)},$ both block-diagonal with respect to the orthogonal decomposition $\Hi_S\oplus\Hi_R$, ensure invariance of $\Hi_S.$ 

By construction, for all $i = 0, \ldots, j$, either $\bigcap_k\ker \bar R_{P,k}^{(i)}=\{0\}$ and $Y^{(i)} = I$, or $\bar R_{P,k}^{(i)}=0$ for all $k$ and $Y^{(i)}$ differs from the identity matrix and has the form \eqref{eq:YwhenNotIdentity} or \eqref{eq:YwhenNotIdentity1}. 

Let us prove that no invariant state can have support on $\bigoplus_{i=1}^{j+1} \Hi_S^{(i)}$ by induction. First consider a state with support on $\Hi^{(j+1)}_{S} = \Hi^{(j)}_{R}$ alone:
$$
\bar \rho =
\left[\begin{array}{c|c}
0 & 0 \\
\hline
0 & \bar \rho_R
\end{array}\right].
$$
If $\bigcap_k\ker \bar R_{P,k}^{(j)}=\{0\}$, then  $\bar \rho$ is mapped by $\sum_k\bar R_k \cdot \bar R_k^\dag$ into a state $\bar\rho'$ with non-trivial support on $\Hi^{(j)}_{S}.$ Being in this case $Y^{(j)} = I,$ $Z^{(j)}$ is block-diagonal with respect to the considered decomposition and we cannot thus get $Z^{(j)}\bar\rho'Z^{(j)\dag}=\bar\rho,$ for any $\bar\rho$ in $\mathfrak{D}(\Hi^{(j)}_R).$

On the other hand, if $\bar R_{P,k}^{(j)}=0\;\forall k$, then $Y^{(j)}$ contains off-diagonal full-rank blocks and maps the state
$$
\bar \rho' =
\left[\begin{array}{c|c}
0 & 0 \\
\hline
0 & \sum_k R_{R,k}^{(j)} \bar \rho_R R_{R,k}^{(j) \dag}
\end{array}\right]
$$
into a state with non-trivial support on $\Hi^{(j)}_{S}$. The subsequent application of $Z^{(j-1)}$ will then map the state into a state with nontrivial support on $\bigoplus_{i=1}^j \Hi_S^{(i)}$, and therefore $\bar \rho$ cannot be invariant.

Let us now proceed with the inductive step, with $m$ as the induction index.
Assume that no invariant state can have support on $\bigoplus_{i=j+1-m}^{j+1}\Hi_S^{(i)}$ alone (induction hypothesis), and consider the subspace $\bigoplus_{i=j-m}^{j+1}\Hi_S^{(i)}.$
By the induction hypothesis if there were an invariant state with support on this subspace, it would be in the form
$$
\bar \rho = 
\left[\begin{array}{c|c|c}
0 & 0 & 0 \\
\hline
0 & \bar \rho_S & \bar \rho_P \\
\hline
0 & \bar \rho_P^\dag & \bar \rho_R 
\end{array}\right]
$$
with $\bar \rho_S \ne 0$ having support on $\Hi_S^{(j-m)}$. Let us rewrite
$$
Z^{(j)} = Z^{(j-m-1)} Y^{(j-m)} Z_{(j-m+1)}
$$
where $Z_{(j-m+1)} = Y^{(j-m+1)} \cdot \ldots \cdot Y^{(j)}$.

Again, we have two cases. If $\bigcap_k\ker \bar R_{P,k}^{(j-m-1)}=\{0\}$, then $Y^{(j-m)}= I$ and $\bar \rho$ is mapped by $\bar R_k$ into a state with non trivial support on $\Hi_S^{(j-m-1)}$. The subsequent application of $Z_{(j-m+1)}$ and of $Y^{(j-m)}$ does not affect this, and because of $Z^{(j-m-1)}$, the first complete iteration will map $\bar \rho$ into a state with non trivial support on $\bigoplus_{i=1}^{j-m-1} \Hi_S^{(i)}$. Therefore $\bar \rho$ cannot be invariant.

On the other hand, if $\bar R_{P,k}^{(j-m-1)}=0\;\forall k$, then $Y^{(j-m)}$ has the form \eqref{eq:YwhenNotIdentity} and the closed loop evolution of $\bar \rho$ is
\beqan
\bar \rho' &=& \sum_k \left(Z^{(j-m-1)} Y^{(j-m)}\right. \underbrace{Z_{(j-m+1)} \bar R_k \bar \rho \bar R_k^\dag Z_{(j-m+1)}^\dag}_{\tilde \rho_k}\\&& \left. Y^{(j-m) \dag} Z^{(j-m-1) \dag}\right).
\eeqan
If $\tilde \rho_k$ has support on $\bigoplus_{i=j+1-m}^{j+1} \Hi_S^{(i)}$ for all $k$, then $\bar \rho'$ will have the same support, and therefore $\bar \rho$ is not invariant. If instead $\tilde \rho_k$ has non trivial support on $\Hi_S^{(j-m)}$ for some $k$, then because of the subsequent application of $Z^{(j-m-1)} Y^{(j-m)}$, $\bar \rho'$ will have non trivial support on $\bigoplus_{i=1}^{j-m-1} \Hi_S^{(i)}$, and again $\bar \rho$ is not invariant.

When the induction process reaches $m=j-1$, then it states that no invariant states are supported on $\Hi_S^{(1)} \oplus \cdots \oplus \Hi_S^{(j)} \oplus \Hi_R^{(j)}$, and therefore according to Theorem \ref{th:attractivity} global asymptotic stability of the subspace $\mathcal S$ is achieved.
\end{IEEEproof}

The algorithm is clearly constructive. We then get the following:

\begin{cor}
A certain subspace $\Hi_S$ can be made GAS if and only if the $R_{P,k}$ blocks of the canonical R-factors, computed with respect to the decomposition $\Hi_I=\Hi_S\oplus \Hi_R$, are not all zero.
\end{cor}


\section{A Toy Problem}

\newcommand{\sqrttwo}{\ensuremath \frac{\sqrt{2}}{2}}
\newcommand{\sqrtfour}{\ensuremath \frac{\sqrt{2}}{4}}

We consider in this example a \emph{two-qubit system}, defined on a Hilbert space $\mathcal H_I \simeq \mathbb C^2 \otimes \mathbb C^2$. Consider the task of stabilizing the maximally entangled state 
\begin{equation}
	\rho_d = \frac12 \left( \ket{00} + \ket{11} \right) \left( \bra{00} + \bra{11} \right) 	
	\label{eq:maxEntangledState}
\end{equation}
which has the following representation in the \emph{computational basis} $\mathcal C = \{\ket{ab} =  \ket{a} \otimes \ket{b} |  a,b = 0,1\}$:
$$
\rho_d = 
\begin{bmatrix}
	1 & 0 & 0 & 1 \\
	0 & 0 & 0 & 0 \\
	0 & 0 & 0 & 0 \\
	1 & 0 & 0 & 1 
\end{bmatrix}.
$$

In order to apply the proposed control design technique, let us consider a different basis $\mathcal B$ such that in the new representation $\rho_d^B = \text{diag}\left(\left[\begin{smallmatrix} 1 & 0 & 0 & 0 \end{smallmatrix}\right]\right)$. This can be achieved by considering the \emph{Bell-basis}
$$
\mathcal B = \left\{
\frac{\ket{00} + \ket{11}}{\sqrt{2}},
\frac{\ket{00} - \ket{11}}{\sqrt{2}},
\frac{\ket{01} + \ket{10}}{\sqrt{2}},
\frac{\ket{01} - \ket{10}}{\sqrt{2}}
\right\}.
$$
Let $B$ be the unitary matrix realizing the change of basis, i.e. $\rho_d^{\mathcal B} = B^\dag \rho_d B$. Consider the space decomposition
$$
\mathcal H_I = \mathcal H_S \oplus \mathcal H_R
$$
where $\mathcal H_S = \text{span}\left\{\frac{1}{\sqrt{2}} (\ket{00} + \ket{11})\right\}$ and $\mathcal H_R = \mathcal H_S^\perp$. We have then successfully casted the problem of stabilizing the maximally entangled state \eqref{eq:maxEntangledState} into the problem of achieving asymptotic stability of the subspace $\mathcal H_S$.
Suppose that the following generalized measurement is available
$$
\mathcal T[\rho ] = \sum_{k=1}^3 M_k^{\phantom{\dag}} \rho M_k^\dag
$$
with operators (represented in the computational basis):
	\beqan
	&&M_1 = \frac{1}{\sqrt{4}} \left( \sigma_+ \otimes I \right) ,\quad
	 M_2 = \frac{1}{\sqrt{4}}  \left( I \otimes \sigma_+ \right),\\
	 && M_3 = \sqrt{I-M_1^\dag M_1-M_2^\dag M_2}.
	\eeqan
where $\sigma_+ = \left[\begin{smallmatrix} 0 & 1 \\ 0 & 0 \end{smallmatrix}\right]$. These Kraus operators may be used to describe a discrete-time spontaneous emission process, where the event associated to $M_{1,2}$ corresponds to the decay of one qubit (with probability $\smallfrac{1}{4}$ each), and we neglect the event of the two qubits decaying in the same time interval.
In the Bell basis, the operators take the form 
\beqan
&& M_1^B = 
	\left[\begin{smallmatrix}
	0			& 0			& \smallfrac{1}{4}		& -\smallfrac{1}{4} \\
	0			& 0			& \smallfrac{1}{4}		& -\smallfrac{1}{4} \\
	\smallfrac{1}{4}	& -\smallfrac{1}{4}	& 0			& 0 \\
	\smallfrac{1}{4}	& -\smallfrac{1}{4}	& 0			& 0 
	\end{smallmatrix}\right],
\quad  M_2^B = 
	\left[\begin{smallmatrix}
	0			& 0			& \smallfrac{1}{4}		& \smallfrac{1}{4} \\
	0			& 0			& \smallfrac{1}{4}		& \smallfrac{1}{4} \\
	\smallfrac{1}{4}	& -\smallfrac{1}{4}	& 0			& 0 \\
	-\smallfrac{1}{4}	& \smallfrac{1}{4}		& 0			& 0 
	\end{smallmatrix}\right],
\\%
&& M_3^B = 
	\left[\begin{smallmatrix}
	0.8536	& 0.1464	& 0			& 0 \\
	0.1464	& 0.8536	& 0			& 0 \\
	0			& 0			& 0.8660	& 0 \\
	0			& 0			& 0	& 0.8660 
	\end{smallmatrix}\right].
\eeqan

Let us then apply the algorithm developed in section \ref{engineering}. The canonical QR decomposition of the matrices $M_k^B$ returns the following triangular factors (we do not report here the corresponding orthogonal matrices $Q_k$, see \eqref{finalU} for the final form of the controls):
\beqan
&& R_1 = 
	\left[\begin{smallmatrix}
	\smallfrac{\sqrt{2}}{4}		& -\smallfrac{\sqrt{2}}{4}	& 0			& 0 \\
	0			& 0			& \smallfrac{\sqrt{2}}{4}	& -\smallfrac{\sqrt{2}}{4} \\
	0			& 0			& 0			& 0 \\
	0			& 0			& 0			& 0 
	\end{smallmatrix}\right],
\quad R_2 = 
	\left[\begin{smallmatrix}
	\smallfrac{\sqrt{2}}{4}		& -\smallfrac{\sqrt{2}}{4}	& 0			& 0 \\
	0			& 0			& \smallfrac{\sqrt{2}}{4}	& \smallfrac{\sqrt{2}}{4} \\
	0			& 0			& 0			& 0 \\
	0			& 0			& 0			& 0 
	\end{smallmatrix}\right],
\\
&& R_3 = 
	\left[\begin{smallmatrix}
	0.8660		& 0.2887	    & 0	     	& 0 \\
	0			& 0.8165		& 0	   	    & 0 \\
	0			& 0			& 0.8660		& 0 \\
	0			& 0			& 0			& 0.8660 
	\end{smallmatrix}\right].
\eeqan

According to the proposed approach, by inspection of the upper triangular factors $R_i$ we can decide about the feasibility of the stabilization task. Indeed, as the blocks $R_{P,k},\,k=1,\dots, 3$ are non-zero blocks, namely
\begin{align*}
R_{P,1} &= 
	\left[\begin{smallmatrix}
		-\smallfrac{\sqrt{2}}{4}	& 0			& 0 
	\end{smallmatrix}\right],
& R_{P,2} &=
	\left[\begin{smallmatrix}
		-\smallfrac{\sqrt{2}}{4} & 0			& 0 
	\end{smallmatrix}\right],
& R_{P,3} &= 
	\left[\begin{smallmatrix}
		0.2887	& 0			& 0
	\end{smallmatrix}\right],
\end{align*}
then the stabilization problem is feasible.

Moreover, notice that at this step no further transformation is needed on the matrices, as the obtained $R$ factors are already decomposed according to
$$
\mathcal H_I = \mathcal H_S \oplus \mathcal H_{S}^{(1)} \oplus \mathcal H_{R}^{(1)}.
$$
where $\mathcal H_R^{(1)} = \bigcap_{k} \ker R_{P,k}$.
Continuing with the iteration, we have then to determine the subspace $\mathcal H_R^{(2)} = \bigcap_{k} \ker R_{P,k}^{(1)}$. By inspection one can see that this space is empty, and therefore the iteration stops successfully.
The set of unitary controls that have to be applied when the corresponding outcome $k$ is measured is then
$$
U_k = B Q_k^\dag B^\dag,
$$
that is:
\beqa\label{finalU}
&& U_1 = 
	\left[\begin{smallmatrix}
	\smallfrac{\sqrt{2}}{2}	& 0	& 0			& -\smallfrac{\sqrt{2}}{2} \\
	\smallfrac{\sqrt{2}}{2}			& 0			& 0			&\smallfrac{\sqrt{2}}{2} \\
	0			& 0			& 1			& 0 \\
	0	& -1	& 0			& 0 \\
	\end{smallmatrix}\right],
\quad U_2 = 
	\left[\begin{smallmatrix}
	\smallfrac{\sqrt{2}}{2} & 0			& 0  & -\smallfrac{\sqrt{2}}{2} \\
	0			& 1			& 0			& 0 \\
	\smallfrac{\sqrt{2}}{2}		& 0			& 0			& \smallfrac{\sqrt{2}}{2} \\
	0	& 0			& -1	& 0 \\
	\end{smallmatrix}\right],
\nonumber \\
&& U_3 = 
	\left[\begin{smallmatrix}
	0.9856 	& 0	& 0			& 0.1691 \\
	0			& 1			& 0			& 0 \\
	0			& 0			& 1			& 0 \\
	-0.1691 & 0	& 0			& 0.9856\\
	\end{smallmatrix}\right].
\eeqa

It can be shown by direct computation that the Hamiltonians needed to implement these unitary transformation (using ideally unbounded control pulses in order to make the dissipation effect negligible on when the control is acting) form a 3-dimensional control algebra \cite{dalessandro-book}.


\section{Conclusions}
Completely positive, trace-preserving maps represent general quantum dynamics for open systems, and if the environment is memoryless, also represent generators of discrete-time quantum Markov semigroups. Theorem \ref{th:attractivity} provides a characterization of the semigroup dynamics that render a certain pure state, or the set of states with support on a subspace, attractive, by employing LaSalle's Invariance Principle. In order to exploit this result for constructive design of stabilizing unitary feedback control strategies, we developed a suitable linear algebraic tool, which holds some interest {\em per se}. We proved that a {\em canonical} QR decomposition can be derived by specializing the well-known orthonormalization approach, and that it is key to study the potential of the feedback control scheme presented in Section \ref{engineering}. In fact, we determined which quantum generalized measurements can be {simulated} controlling a given one, and which pure states or subspaces can be rendered globally asymptotically stable. Theorem \ref{th:algorithm} gives a constructive procedure to build the controls, and also a simple test on the existence of such controls: If the algorithm does not stop on the first step, then the control problem has a solution.
We believe that the provided results also represent a mathematical standpoint from which interesting, and more challenging, control problems can be tackled, in particular when the control choice is constrained by a multipartite structure of the system of interest.

\appendix

\subsection{Quantum Measurements}\label{measurements}
\label{app:QuantumMeasurements}


%

\subsubsection{Projective Measurements}
In quantum mechanics, observable quantities are associated to Hermitian operators, with their spectrum associated to the possible outcomes. Suppose that we are interested in measuring the {\em observable}
$C=\sum_{i} c_i \Pi_i$.
The basic postulates that describe the quantum (strong, projective, or von Neumann's) measurements are the following:

\begin{itemize}
   \item [(i')] The probability of obtaining $c_i$ as the outcome
   of a measure on a system described by the density operator $\rho$
   is $p_i=\Tr(\rho\Pi_i).$ 
   \item [(ii')] [L\"uders's Postulate] Immediately after a measurement
   that gives $c_i$ as an outcome the system state
   becomes: $\rho|_i=\frac{1}{\trace(\Pi_i\rho\Pi_i)}\Pi_i\rho\Pi_i.$
\end{itemize}

Notice that the spectrum of the observable does not play any role in the computation of the probabilities.

\subsubsection{Generalized Measurements}

If we get information about a quantum system by measuring another system which is correlated to the former, the projective measurement formalism is not enough, but it can be used to derive a more general one. A typical procedure to obtain generalized measurements on a quantum system of interest is the following:
\begin{itemize}
\item The system of interest ${\cal A}$ is augmented by adding another subsystem ${\cal B}$, initially decoupled from ${\cal A}$. Let $\rho_{\cal A}\otimes \rho_{\cal B},$  with $\rho_{\cal B}=\ket{\phi}\bra{\phi},$ be the joint state;
\item The two systems are coupled through a joint unitary evolution $U_{\cal A B}$;
\item A direct, von Neumann measurement of an observable $X_{\cal B}=\sum_jx_j\Pi_j,$ $\Pi_j=\ket{\xi_j}\bra{\xi_j},$  is performed on ${\cal B}.$ 
\item The conditioned state of the joint system after the measurement is then of the form $$\rho_{\cal AB}|_j=\frac{1}{p_j}(I_{\cal A}\otimes\Pi_j)U_{\cal A B}(\rho_{\cal A}\otimes \rho_{\cal B})U_{\cal A B}^\dag(I_{\cal A}\otimes\Pi_j)=\rho_{{\cal A},j}'\otimes\Pi_j,$$
with $p_j$ the probability of obtaining the $j$-th outcome.
\item One can compute the effect of the measurement on ${\cal A}$ alone, which is nontrivial if $U_{\cal A B}$ entangled the two subsystems, i.e. $U_{\cal A B}(\rho_{\cal A}\otimes \rho_{\cal B})U_{\cal AB}^\dag$ cannot be written in factorized form.
One then gets that $\rho_{{\cal A},j}'=\frac{1}{p_j}M_j\rho_{\cal A} M_j^\dag,$ with $M_j=\bra{\xi_j}U_{\cal A B}\ket{\phi}$.
\end{itemize}
If now the average over the possible outcomes is taken, we obtain a state transformation in Kraus form. This construction is actually general, in the sense that if the dimension of ${\cal B}$ corresponds (at least) to the necessary number of outcomes, {\em any} Kraus map can be actually generated this way.

\bibliography{bibliography}
\bibliographystyle{plain}

\end{document}